\documentclass[longauth, letter, traditabstract]{aa} 
\usepackage{graphicx}
\usepackage{natbib}
\bibpunct{(}{)}{,}{a}{}{;}
\usepackage{amssymb}
\usepackage{amsmath}
\usepackage{url}
\usepackage{multirow}
\usepackage{txfonts}
\begin{document}
   \title{Water in low-mass star-forming regions with \textit{Herschel}\thanks{\textit{Herschel} is an ESA space observatory with science instruments provided by European-led Principal Investigator consortia and with important participation from NASA.}}
\subtitle{HIFI spectroscopy of NGC1333}


\author{
	 L.E.~Kristensen\inst{1}
\and R.~Visser\inst{1}
\and E.F.~van~Dishoeck\inst{1,2}
\and U.A.~Y{\i}ld{\i}z\inst{1}
\and S.D.~Doty\inst{3}
\and G.J.~Herczeg\inst{2}
\and F.-C. Liu\inst{4}
\and B. Parise\inst{4}
\and J.K.~J{\o}rgensen\inst{5}
\and T.A.~van~Kempen\inst{6}
\and C.~Brinch\inst{1}
\and S.F.~Wampfler\inst{7}
\and S.~Bruderer\inst{7}
\and A.O.~Benz\inst{7}
\and M.R.~Hogerheijde\inst{1}
\and E. Deul\inst{1}
\and R.~Bachiller\inst{8}
\and A.~Baudry\inst{9}
\and M.~Benedettini\inst{10}
\and E.A.~Bergin\inst{11}
\and P.~Bjerkeli\inst{12}
\and G.A.~Blake\inst{13}
\and S.~Bontemps\inst{9}
\and J.~Braine\inst{9}
\and P.~Caselli\inst{14,15}
\and J.~Cernicharo\inst{16}
\and C.~Codella\inst{15}
\and F.~Daniel\inst{16}
\and Th.~de~Graauw\inst{17}
\and A.M.~di~Giorgio\inst{10}
\and C.~Dominik\inst{18,19}
\and P.~Encrenaz\inst{20}
\and M.~Fich\inst{21}
\and A.~Fuente\inst{22}
\and T.~Giannini\inst{23}
\and J.R.~Goicoechea\inst{16}
\and F.~Helmich\inst{17}
\and F.~Herpin\inst{9}
\and T.~Jacq\inst{9}
\and D.~Johnstone\inst{24,25}
\and M.J.~Kaufman\inst{26}
\and B.~Larsson\inst{27}
\and D.~Lis\inst{28}
\and R.~Liseau\inst{12}
\and M.~Marseille\inst{17}
\and C.~M$^{\rm c}$Coey\inst{21,29}
\and G.~Melnick\inst{6}
\and D.~Neufeld\inst{30}
\and B.~Nisini\inst{23}
\and M.~Olberg\inst{12}
\and J.C.~Pearson\inst{31}
\and R.~Plume\inst{32}
\and C.~Risacher\inst{17}
\and J.~Santiago-Garc\'{i}a\inst{33}
\and P.~Saraceno\inst{10}
\and R.~Shipman\inst{17}
\and M.~Tafalla\inst{8}
\and A.G.G.M.~Tielens\inst{1}
\and F.~van der Tak\inst{17,34}
\and F.~Wyrowski\inst{4}
\and D.~Beintema\inst{17}
\and A.~de Jonge\inst{17}
\and P.~Dieleman\inst{17}
\and V.~Ossenkopf\inst{35}
\and P.~Roelfsema\inst{17}
\and J.~Stutzki\inst{35}
\and N.~Whyborn\inst{36}
}

\institute{
Leiden Observatory, Leiden University, PO Box 9513, 2300 RA Leiden, The Netherlands
\and
Max Planck Institut f\"{u}r Extraterrestrische Physik, Giessenbachstrasse 1, 85748 Garching, Germany
\and
Department of Physics and Astronomy, Denison University, Granville, OH, 43023, USA
\and
Max-Planck-Institut f\"{u}r Radioastronomie, Auf dem H\"{u}gel 69, 53121 Bonn, Germany
\and
Centre for Star and Planet Formation, Natural History Museum of Denmark, University of Copenhagen,
{\O}ster Voldgade 5-7, DK-1350 Copenhagen K., Denmark
\and
Harvard-Smithsonian Center for Astrophysics, 60 Garden Street, MS 42, Cambridge, MA 02138, USA
\and
Institute of Astronomy, ETH Zurich, 8093 Zurich, Switzerland
\and
Observatorio Astron\'{o}mico Nacional (IGN), Calle Alfonso XII,3. 28014, Madrid, Spain
\and
Universit\'{e} de Bordeaux, Laboratoire d'Astrophysique de Bordeaux, France; CNRS/INSU, UMR 5804, Floirac, France
\and
INAF - Instituto di Fisica dello Spazio Interplanetario, Area di Ricerca di Tor Vergata, via Fosso del Cavaliere 100, 00133 Roma, Italy
\and
Department of Astronomy, University of Michigan, 500 Church Street, Ann Arbor, MI 48109-1042, USA
\and
Department of Radio and Space Science, Chalmers University of Technology, Onsala Space Observatory, 439 92 Onsala, Sweden
\and
California Institute of Technology, Division of Geological and Planetary Sciences, MS 150-21, Pasadena, CA 91125, USA
\and
School of Physics and Astronomy, University of Leeds, Leeds LS2 9JT, UK
\and
INAF - Osservatorio Astrofisico di Arcetri, Largo E. Fermi 5, 50125 Firenze, Italy
\and
Centro de Astrobiolog{\'i}a, Departamento de Astrof{\'i}sica, CSIC-INTA, Carretera de Ajalvir, Km 4, Torrej{\'o}n de Ardoz. 28850, Madrid, Spain
\and
SRON Netherlands Institute for Space Research, PO Box 800, 9700 AV, Groningen, The Netherlands
\and
Astronomical Institute Anton Pannekoek, University of Amsterdam, Kruislaan 403, 1098 SJ Amsterdam, The Netherlands
\and
Department of Astrophysics/IMAPP, Radboud University Nijmegen, P.O. Box 9010, 6500 GL Nijmegen, The Netherlands
\and
LERMA and UMR 8112 du CNRS, Observatoire de Paris, 61 Av. de l'Observatoire, 75014 Paris, France
\and
University of Waterloo, Department of Physics and Astronomy, Waterloo, Ontario, Canada
\and
Observatorio Astron\'{o}mico Nacional, Apartado 112, 28803 Alcal\'{a} de Henares, Spain
\and
INAF - Osservatorio Astronomico di Roma, 00040 Monte Porzio catone, Italy
\and
National Research Council Canada, Herzberg Institute of Astrophysics, 5071 West Saanich Road, Victoria, BC V9E 2E7, Canada
\and
Department of Physics and Astronomy, University of Victoria, Victoria, BC V8P 1A1, Canada
\and
Department of Physics and Astronomy, San Jose State University, One Washington Square, San Jose, CA 95192, USA
\and
Department of Astronomy, Stockholm University, AlbaNova, 106 91 Stockholm, Sweden
\and
California Institute of Technology, Cahill Center for Astronomy and Astrophysics, MS 301-17, Pasadena, CA 91125, USA
\and
University of Western Ontario, Department of Physics \& Astronomy, London, Ontario, Canada N6A 3K7 
\and
Department of Physics and Astronomy, Johns Hopkins University, 3400 North Charles Street, Baltimore, MD 21218, USA
\and
Jet Propulsion Laboratory, California Institute of Technology, Pasadena, CA 91109, USA
\and
Department of Physics and Astronomy, University of Calgary, Calgary, T2N 1N4, AB, Canada
\and
Instituto de RadioAstronom\'{i}a Milim\'{e}trica, Avenida Divina Pastora, 7, N\'{u}cleo Central E 18012 Granada, Spain
\and
Kapteyn Astronomical Institute, University of Groningen, PO Box 800, 9700 AV, Groningen, The Netherlands
\and
KOSMA, I. Physik. Institut, Universit{\"a}t zu K{\"o}ln, Z{\"u}lpicher Str. 77, D 50937 K{\"o}ln, Germany
\and
Atacama Large Millimeter/Submillimeter Array, Joint ALMA Office, Santiago, Chile 
}

\date{Draft: \today}

\titlerunning{\textit{Herschel}-HIFI spectroscopy of NGC1333}


\abstract
{`Water In Star-forming regions with \textit{Herschel}' (WISH) is a key programme dedicated to studying the role of water and related species during the star-formation process and constraining the physical and chemical properties of young stellar objects. The Heterodyne Instrument for the Far-Infrared (HIFI) on the \textit{Herschel} Space Observatory observed three deeply embedded protostars in the low-mass star-forming region NGC1333 in several H$_2^{16}$O, H$_2^{18}$O, and CO transitions. Line profiles are resolved for five H$_2^{16}$O transitions in each source, revealing them to be surprisingly complex. The line profiles are decomposed into broad ($>$20 km\,s$^{-1}$), medium-broad ($\sim$5--10 km\,s$^{-1}$), and narrow ($<$5 km\,s$^{-1}$) components. The H$_2^{18}$O emission is only detected in broad 1$_{10}$--1$_{01}$ lines ($>$20 km\,s$^{-1}$), indicating that its physical origin is the same as for the broad H$_2^{16}$O component. In one of the sources, IRAS4A, an inverse P Cygni profile is observed, a clear sign of infall in the envelope. From the line profiles alone, it is clear that the bulk of emission arises from shocks, both on small ($\lesssim$1000 AU) and large scales along the outflow cavity walls ($\sim$10\,000 AU). The H$_2$O line profiles are compared to CO line profiles to constrain the H$_2$O abundance as a function of velocity within these shocked regions. The H$_2$O/CO abundance ratios are measured to be in the range of $\sim$0.1--1, corresponding to H$_2$O abundances of $\sim$10$^{-5}$--10$^{-4}$ with respect to H$_2$. Approximately 5--10\% of the gas is hot enough for all oxygen to be driven into water in warm post-shock gas, mostly at high velocities.}


\maketitle

\section{Introduction}

In the deeply embedded phase of low-mass star formation, it is often only possible to trace the dynamics of gas in a young stellar object (YSO) by analysing resolved emission-line profiles. The various dynamical processes include infall from the surrounding envelope towards the central protostar, molecular outflows caused by jets ejected from the central object, and strong turbulence induced within the inner parts of the envelope by small-scale shocks \citep{arce07, jorgensen07}. One of the goals of the Water In Star-forming regions with \textit{Herschel} (WISH) key programme is to use water as a probe of these processes and determine its abundance in the various components as a function of evolution (van Dishoeck et al. in prep.).

Spectrally resolved observations of the H$_2$O 1$_{10}$--1$_{01}$ line at 557 GHz with ODIN and SWAS towards low-mass star-forming regions have revealed it to be broad, $\sim$20 km\,s$^{-1}$, indicative of an origin in shocks \citep[e.g.,][]{bergin03}. Within the large beams (2\arcmin\ and 4\arcmin), where both the envelope and the entire outflow are present, outflow emission most likely dominates. Observations and subsequent modelling of the more highly excited H$_2$O lines with ISO-LWS were unable to distinguish between an origin in shocks or an infalling envelope \citep[e.g.,][]{ceccarelli96, Nisini02, Maret02}. \textit{Herschel}-HIFI has a much higher sensitivity, higher spectral resolution, and smaller beam than previous space-based missions, thus is perfectly suited to addressing this question. Complementary CO data presented by \citet{Yildiz10} are used to constrain the role of the envelope and  determine outflow temperatures and densities.

NGC1333 is a well-studied region of clustered, low-mass star formation at a distance of 235 pc \citep{hirota08}.  In particular, the three deeply embedded, low-mass class 0 objects IRAS2A, IRAS4A, and IRAS4B have been observed extensively with ground-based submillimetre telescopes \citep[e.g.,][]{jorgensen05, maret05} and interferometers \citep[e.g.,][]{difrancesco01, jorgensen07}. All sources have strong outflows extending over arcmin scales ($>$15\,000 AU). Both IRAS4A and 4B consist of multiple protostars \citep[e.g.,][]{choi05}. Because of the similarities between the three sources in terms of luminosity (20, 5.8, and 3.8 $L_\odot$), envelope mass \citep[1.0, 4.5, and 2.9 $M_\odot$;][]{jorgensen09} and presumably also age, they provide ideal grounds for comparing YSOs in the same region.

\section{Observations and results}
\label{sec:obs}

\begin{figure}
\begin{center}
\includegraphics[width=\columnwidth, angle=0]{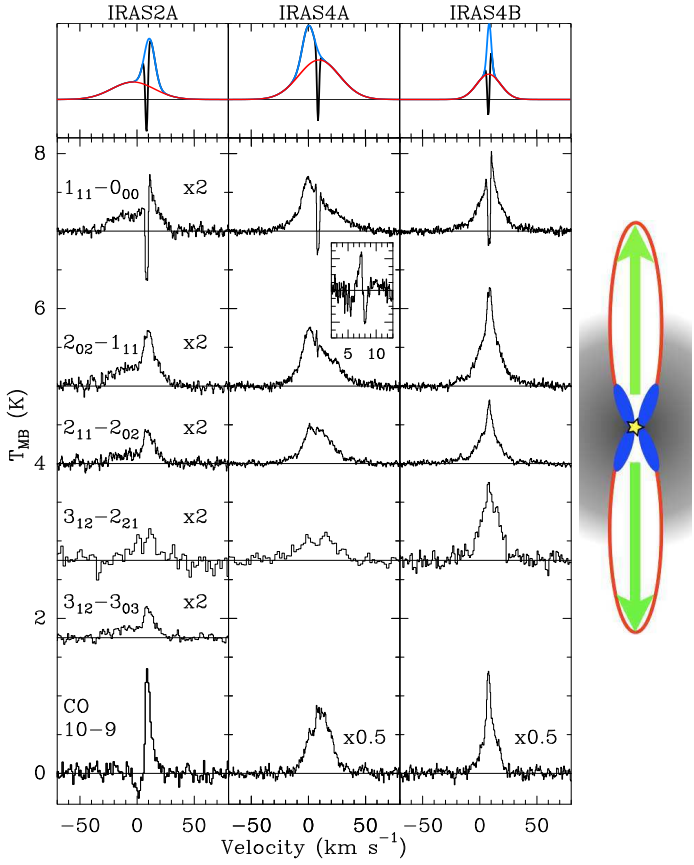}
\end{center}
\caption{H$_2$O spectra of the three NGC1333 sources. CO 10--9 is shown for comparison \citep{Yildiz10}; the CO 10--9 emission in IRAS2A is affected by chopping into outflow material. The top panel shows the decomposition into broad (red), medium (blue), and narrow (black) components. The cartoon illustrates the physical origin of each component. The inset shows a zoom on the inverse P Cygni profile in the H$_2$O 2$_{02}$--1$_{11}$ line of IRAS4A, where the other components have been subtracted; the vertical scale ranges from $-$0.3 to 0.3 K.}
\label{fig:spectra}
\end{figure}

\begin{figure}
\begin{center}
\includegraphics[width=.37\columnwidth,angle=270]{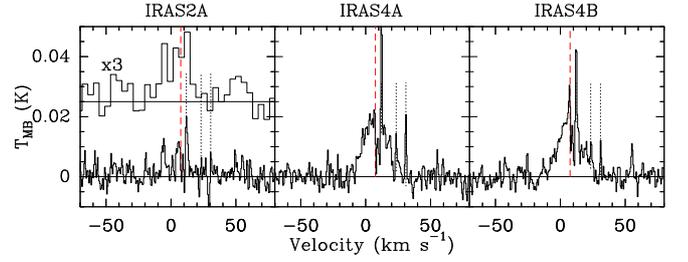}
\end{center}
\caption{H$_2^{18}$O 1$_{10}$--1$_{01}$ spectra of the three NGC1333 sources along with the CH 536 GHz triplet from the lower sideband (dotted lines). Spectra are shown for a channel size of 0.25 km\,s$^{-1}$. The spectrum of IRAS2A has been rebinned to 4 km\,s$^{-1}$ to illustrate the detection of a broad feature. The  red line shows the source velocity at $\varv_{\rm LSR}$$\sim$$+$7.5 km\,s$^{-1}$.}
\label{fig:spectra_h218o}
\end{figure}

Three sources in NGC1333, IRAS2A, IRAS4A, and IRAS4B, were observed with HIFI \citep{degraauw10} on \textit{Herschel} \citep{pilbratt10} on March 3--15, 2010 in dual beam switch mode in bands 1, 3, 4, and 5 with a nod of 3\arcmin. Observations detected several transitions of H$_2$O and H$_2^{18}$O in the range $E_{\rm u}/k_{\rm B}$$\approx$50--250 K (Table \ref{tab:h2o_line} in the online appendix). Diffraction-limited beam sizes were in the range 19--40\arcsec\ (4500--9500 AU). In general, the calibration is expected to be accurate to $\sim$20\% and the pointing to $\sim$2\arcsec. Data were reduced with HIPE 3.0. A main-beam efficiency of 0.74 was used throughout. Subsequent analysis was performed in CLASS. The rms was in the range 3--150 mK in 0.5 km\,s$^{-1}$ bins. Linear baselines were subtracted from all spectra, except around 750 GHz (corresponding to the H$_2$O 2$_{11}$--2$_{02}$ transition) where higher-order polynomials are required. A difference in rms was always seen between the H- and V-polarizations, with the rms in the H-polarization being lower. In cases where the difference exceeded 30\% and qualitative differences appear in the line profile, the V-polarization was discarded, otherwise the spectra were averaged.

All targeted lines of H$_2^{16}$O were detected and are listed in Table \ref{tab:h2o_obs} and Fig. \ref{fig:spectra}. The 1$_{10}$--1$_{01}$ transition at 557 GHz was not observed before the sources moved out of visibility. The H$_2^{18}$O 1$_{10}$--1$_{01}$ line was detected in all sources (Fig. \ref{fig:spectra_h218o}), although the detection in IRAS2A was weak ($\sim$5$\sigma$ = 0.13 K\,km\,s$^{-1}$). This line is superposed on the ground-state CH triplet at 536 GHz, observed in the lower sideband (Fig. \ref{fig:spectra_h218o}). Neither the H$_2^{18}$O 1$_{11}$--0$_{00}$ nor the 2$_{02}$--1$_{11}$ line in IRAS2A is detected down to $\sigma$$<$0.06 K\,km\,s$^{-1}$. 

The H$_2$O lines exhibit multiple components: a broad emission component (FWHM$>$$20$ km\,s$^{-1}$) sometimes offset from the source velocity ($\varv_{\rm LSR}$=$+$7.2--7.7 km\,s$^{-1}$); a medium-broad emission component (FWHM$\sim$5--10 km\,s$^{-1}$); and a deep, narrow absorption component (FWHM$\sim$2 km\,s$^{-1}$) seen at the source velocity. The individual components are all reproduced well by Gaussian functions. The absorption is only seen in the H$_2$O 1$_{11}$--0$_{00}$ line and is saturated in IRAS2A and IRAS4A. In IRAS4B, the absorption extends below the continuum level, but is not saturated. Furthermore, the IRAS4A spectrum of the 2$_{02}$--1$_{11}$ line exhibits an inverse P Cygni profile. The shape of the lines is the same within a source; only the relative contribution between the broad and medium components changes. For example, in IRAS2A the ratio of the peak intensities is $\sim$2, independent of the line, whereas in IRAS4A it ranges from 1 to 2. The H$_2^{18}$O line profiles compare well to the broad component seen in H$_2$O, i.e., similar FWHM$>$20 km\,s$^{-1}$ and velocity offset. The width is much larger than isotopologue emission of, e.g., C$^{18}$O ($\sim$1--2 km\,s$^{-1}$) and is centred on the source velocity \citep{Yildiz10}. The medium and narrow components are not seen in the H$_2^{18}$O 1$_{10}$--1$_{01}$ spectra down to an rms of 2--3 mK in 0.5 km\,s$^{-1}$ bins.

\begin{table}
\caption{H$_2$O and H$_2^{18}$O emission\tablefootmark{a} in the NGC1333 sources\tablefootmark{b}.}
\scriptsize
\begin{center}
\begin{tabular}{l r c c c c @{} c c}
\hline \hline
       &            &      & \multicolumn{2}{c}{Medium} & & \multicolumn{2}{c}{Broad} \vspace{1pt}\\ \cline{4-5} \cline{7-8}
Source & Transition & rms\tablefootmark{c}  & $T_{\rm MB}^{\rm peak}$ & $\int T_{\rm MB}~{\rm d}\varv$ & & $T_{\rm MB}^{\rm peak}$ & $\int T_{\rm MB}~{\rm d}\varv$ \\
       &            & (mK) & (K) & (K\,km\,s$^{-1}$) & & (K) & (K\,km\,s$^{-1}$) \\
\hline
IRAS2A  & H$_2$O 1$_{11}$--0$_{00}$ & 22 & 0.22 & 2.7 & & 0.10 & 4.1 \\
        &        2$_{02}$--1$_{11}$ & 25 & 0.25 & 2.5 & & 0.12 & 5.6 \\
        &        2$_{11}$--2$_{02}$ & 23 & 0.16 & 1.9 & & 0.06 & 2.7 \\
        &        3$_{12}$--3$_{03}$ & 79 & 0.14 & 1.5 & & 0.06 & 2.7 \\
        &        3$_{12}$--2$_{21}$ & 52 & 0.13 & 1.5 & & 0.08 & 3.8 \\
        & H$_2^{18}$O 1$_{10}$--1$_{01}$ &  2 & \ldots & $<$0.01 & &   0.01 & 0.14 \\
        &             1$_{11}$--0$_{00}$ & 22 & \ldots & $<$0.07 & & \ldots & $<$0.12 \\
        &             2$_{02}$--1$_{11}$ & 12 & \ldots & $<$0.04 & & \ldots & $<$0.06 \\
        &             3$_{12}$--3$_{03}$ & 19 & \ldots & $<$0.06 & & \ldots & $<$0.10 \\
\multicolumn{3}{l}{FWHM\tablefootmark{d} (km\,s$^{-1}$)} & \multicolumn{2}{c}{10.7$\pm$0.8} & & \multicolumn{2}{c}{42$\pm$3} \\
\multicolumn{3}{l}{$\varv_{\rm LSR}$\tablefootmark{d} (km\,s$^{-1}$)} & \multicolumn{2}{c}{$+$10.7$\pm$0.8} & & \multicolumn{2}{c}{$-$2.3$\pm$3.4} \\ \hline
IRAS4A & H$_2$O 1$_{11}$--0$_{00}$ &  23 & 0.36 & 4.9 & & 0.35 & 15.3 \\
       &        2$_{02}$--1$_{11}$ &  24 & 0.34 & 3.8 & & 0.45 & 18.0 \\
       &        2$_{11}$--2$_{02}$ &  23 & 0.14 & 0.8 & & 0.43 & 14.0 \\
       &        3$_{12}$--2$_{21}$ & 100 & 0.07 & 0.4 & & 0.32 & 13.6 \\
       & H$_2^{18}$O 1$_{10}$--1$_{01}$ &  3 & \ldots & $<$0.01 & &   0.02 & 0.43 \\
       &             1$_{11}$--0$_{00}$ & 23 & \ldots & $<$0.06 & & \ldots & $<$0.09 \\
\multicolumn{3}{l}{FWHM\tablefootmark{d} (km\,s$^{-1}$)} & \multicolumn{2}{c}{11.1$\pm$2.3} & & \multicolumn{2}{c}{37$\pm$4} \\
\multicolumn{3}{l}{$\varv_{\rm LSR}$\tablefootmark{d} (km\,s$^{-1}$)} & \multicolumn{2}{c}{$-$0.6$\pm$0.5} & & \multicolumn{2}{c}{$+$8.7$\pm$1.0} \\ \hline
IRAS4B & H$_2$O 1$_{11}$--0$_{00}$ &  29 &  1.1 &  5.1 & & 0.54 & 14.0 \\
       &        2$_{02}$--1$_{11}$ &  23 & 0.63 &  3.5 & & 0.65 & 17.6 \\
       &        2$_{11}$--2$_{02}$ &  17 & 0.37 &  1.8 & & 0.40 & 10.2 \\
       &        3$_{12}$--2$_{21}$ & 150 & 0.32 &  1.3 & & 0.76 & 17.5 \\
       & H$_2^{18}$O 1$_{10}$--1$_{01}$ &  3 & \ldots & $<$0.01 & &   0.02 & 0.43 \\
       &             1$_{11}$--0$_{00}$ & 16 & \ldots & $<$0.03 & & \ldots & $<$0.04 \\
\multicolumn{3}{l}{FWHM\tablefootmark{d} (km\,s$^{-1}$)} & \multicolumn{2}{c}{4.6$\pm$0.5} & & \multicolumn{2}{c}{24$\pm$2} \\
\multicolumn{3}{l}{$\varv_{\rm LSR}$\tablefootmark{d} (km\,s$^{-1}$)} & \multicolumn{2}{c}{$+$8.1$\pm$0.3} & & \multicolumn{2}{c}{$+$8.0$\pm$0.5} \\
\hline\\
\end{tabular}
\vspace{-8pt}
\tablefoot{
	\tablefoottext{a}{Obtained from Gaussian fits to each component. In the case of H$_2$O 1$_{11}$--0$_{00}$, this includes extrapolation over the absorption feature.}
	\tablefoottext{b}{The coordinates used are for IRAS2A: 03:28:55.6; +31:14:37.1, IRAS4A: 03:29:10.5; +31:13:30.9, IRAS4B: +03:29:12.0; +31:13:08.1 (J2000).}
	\tablefoottext{d}{Measured in 0.5 km\,s$^{-1}$ bins.}
	\tablefoottext{e}{Intensity-weighted average of values determined from Gaussian fits of H$_2^{16}$O emission lines. Uncertainties include statistical errors only.}}
\end{center}
\label{tab:h2o_obs}
\end{table}

\onltab{2}{
\begin{table}
\caption{Observed H$_2$O, H$_2^{18}$O and CH transitions\tablefootmark{a}.}
\scriptsize
\begin{center}
\begin{tabular}{r c c c c c c}
\hline \hline
\multicolumn{1}{c}{Transition} & $\nu$ & $\lambda$   & $E_{\rm u}/k_{\rm B}$ & $A$  & Beam & $t_{\rm int}$\tablefootmark{b} \\
                               & (GHz) & ($\mu$m)    & (K)   & ($10^{-3}$ s$^{-1}$) & ($''$) & (min.) \\
\hline
H$_2$O ~~~1$_{11}$--0$_{00}$ &           1113.34 & 269.27 & \phantom{1}53.4 &           18.42 & 19 & 43.5 \\
2$_{02}$--1$_{11}$           & \phantom{1}987.93 & 303.46 & 100.8           & \phantom{1}5.84 & 22 & 23.3 \\
2$_{11}$--2$_{02}$           & \phantom{1}752.03 & 398.64 & 136.9           & \phantom{1}7.06 & 29 & 18.4 \\
3$_{12}$--3$_{03}$           &           1097.37 & 273.19 & 249.4           &           16.48 & 20 & 32.4 \\
3$_{12}$--2$_{21}$           &           1153.13 & 259.98 & 249.4           & \phantom{1}2.63 & 19 & 13.0 \\
\hline
H$_2^{18}$O ~~~1$_{10}$--1$_{01}$\phantom{\tablefootmark{c}} & \phantom{1}547.68 & 547.39 & \phantom{1}60.5 & \phantom{1}3.59 & 39 & 64.3 \\
1$_{11}$--0$_{00}$\tablefootmark{c} &           1101.70 & 272.12 & \phantom{1}52.9 &           21.27 & 20 & 43.5 \\
2$_{02}$--1$_{11}$\phantom{\tablefootmark{c}} & \phantom{1}994.68 & 301.40 & 100.7 &           \phantom{1}7.05 & 22 & 46.7 \\
3$_{12}$--3$_{03}$\tablefootmark{c} &           1095.16 & 273.74 & 289.7 &                     22.12 & 20 & 32.4 \\
\hline
CH\tablefootmark{d} ~~~3/2,2$^-$--1/2,1$^+$ & \phantom{1}536.76 & 558.52 & \phantom{1}25.8 & \phantom{1}0.66 & 39 & \\
   3/2,1$^-$--1/2,1$^+$ & \phantom{1}536.78 & 558.50 & \phantom{1}25.8 & \phantom{1}0.23 & 39 & \\
   3/2,1$^-$--1/2,0$^+$ & \phantom{1}536.80 & 558.48 & \phantom{1}25.8 & \phantom{1}0.46 & 39 &\\
\hline
\end{tabular}
\tablefoot{
	\tablefoottext{a}{From the JPL database of molecular spectroscopy \citep{pickett98}.}
	\tablefoottext{b}{Total on $+$ off integration time.}
	\tablefoottext{c}{Observed in the same setting as the main isotopologue.}
	\tablefoottext{d}{Observed with H$_2^{18}$O 1$_{10}$--1$_{01}$.}}
\end{center}
\label{tab:h2o_line}
\end{table}
}

The upper limits to the H$_2^{18}$O 1$_{11}$--0$_{00}$ line are invaluable for estimating upper limits to the optical depth, $\tau$. In the following, the limit on $\tau$ is derived for the integrated intensity; in the line wings, $\tau$ is most likely lower \citep{Yildiz10}. In the broad component, the limit ranges from 0.4 (IRAS4B) to 2 (IRAS2A), whereas it ranges from 1.1 (IRAS4B) to 2.7 (IRAS2A) for the medium component of the H$_2^{16}$O 1$_{11}$--0$_{00}$ line. Performing the same analysis to the upper limit on the H$_2^{18}$O 2$_{02}$--1$_{11}$ line observed in IRAS2A, infers an upper limit to the optical depth of H$_2^{16}$O 2$_{02}$--1$_{11}$ of 1.5 for the medium component and 1.9 for the broad. Thus it is likely that neither the broad nor the medium components are very optically thick.

\section{Discussion}
\label{sec:disc}

Many physical components in a YSO are directly traced by the line profiles presented here, including the infalling envelope and shocks along the cavity walls. In the following, each component is discussed in detail, and the H$_2$O abundance is estimated in the various physical components.

\subsection{Line profiles}

The most prominent feature of all the observed line profiles is their width. All line wings span a range of velocities of $\sim$40--70 km\,s$^{-1}$ at their base. The width alone indicates that the bulk of the H$_2$O emission originates in shocks along the cavity walls, also called shells, seen traditionally as the standard high-velocity component in CO outflow data, but with broader line-widths due to water enhancement at higher velocities \citep[Sect. \ref{sect:abun}][]{bachiller90, santiago-garcia09}. The shocks release water from the grains by means of sputtering and in high-temperature regions all free oxygen is driven into water. The shocked regions may be illuminated by FUV radiation originating in the star-disk boundary layer, thus further enhancing the water abundance by means of photodesorption. The broad emission seen in the H$^{18}_2$O 1$_{10}$--1$_{01}$ line arises in the same shocks (see cartoon in Fig. \ref{fig:spectra}).

\begin{figure}
\begin{center}
\includegraphics[width=.45\columnwidth]{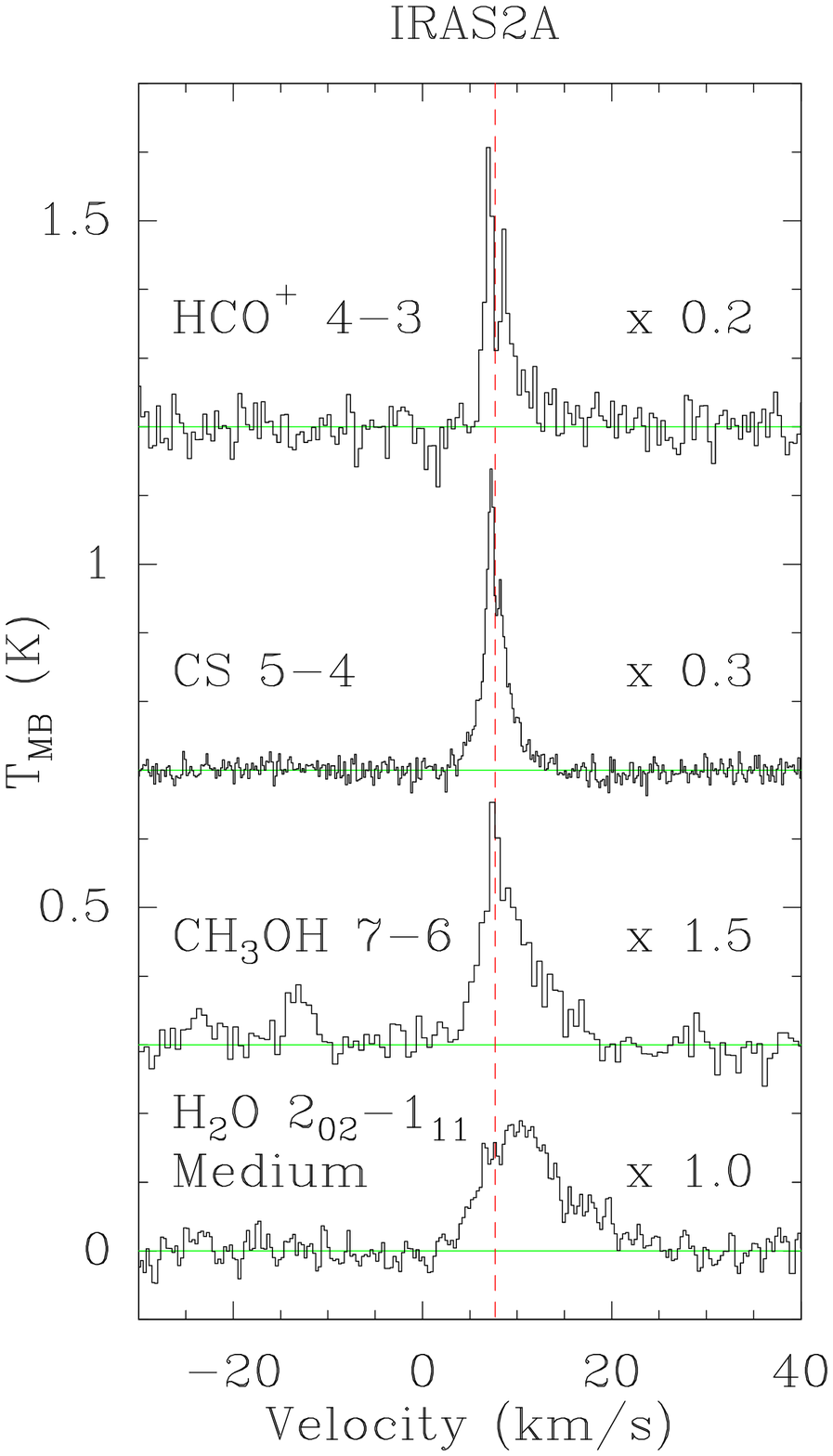}
\includegraphics[width=.45\columnwidth]{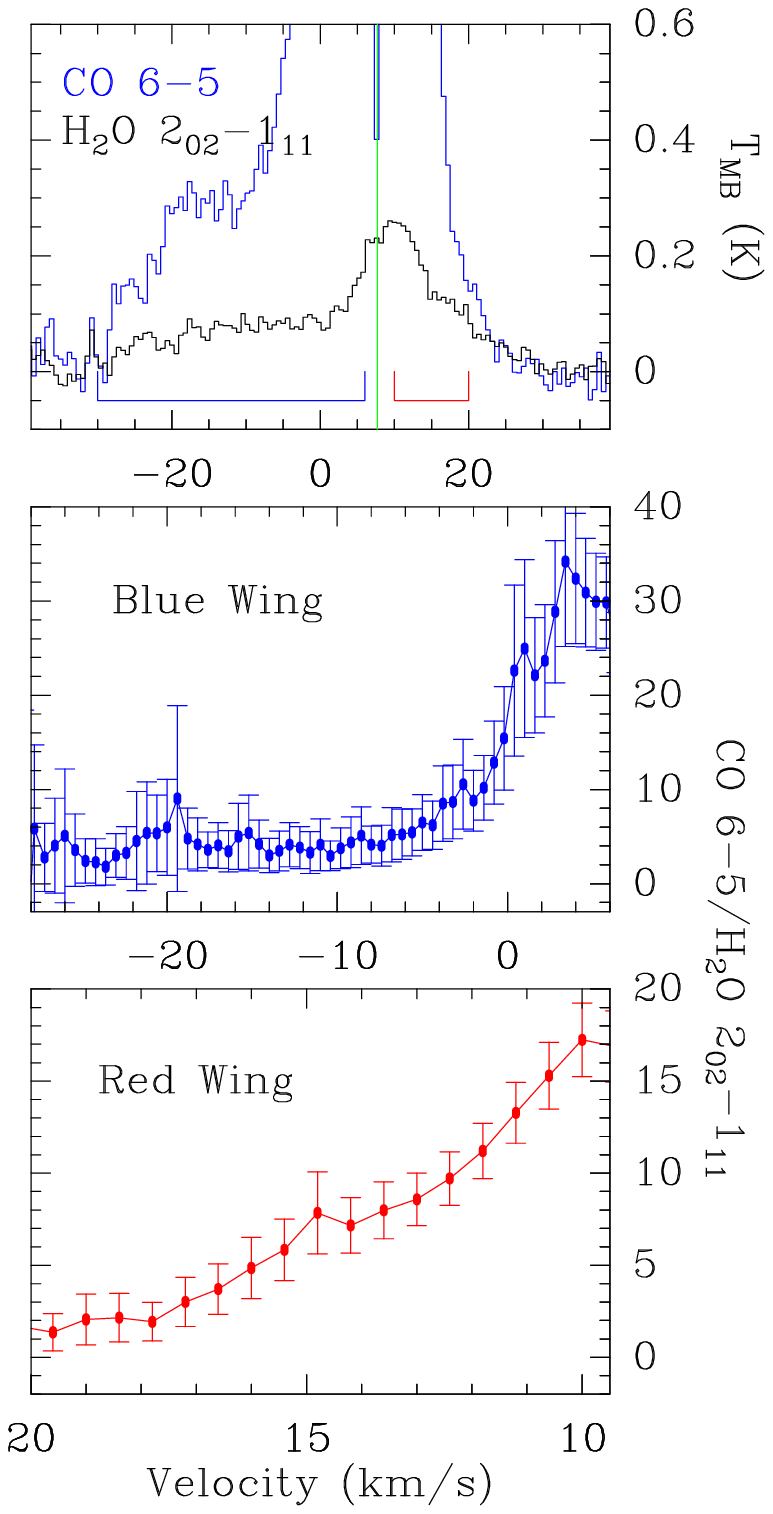}
\end{center}
\caption{{\it Left.} Comparison between the medium component in IRAS2A and other species observed with ground-based telescopes. The broad component has been subtracted for easy comparison. The vertical red line indicates the source velocity at $+$7.7 km\,s$^{-1}$. {\it Right.} Comparison between H$_2$O 2$_{02}$--1$_{11}$ and CO 6--5 obtained with APEX-CHAMP$^+$, and emission ratios for the blue- and red-shifted outflow lobes.}
\label{fig:i2a_co65}
\end{figure}

The medium components (FWHM$\sim$5--10 km\,s$^{-1}$) are most likely also caused by shocks, although presumably on a smaller spatial scale and in denser material than the shocks discussed above. For example, the medium component in IRAS2A is seen in other grain-product species such as CH$_3$OH \citep[][Fig. \ref{fig:i2a_co65}]{jorgensen05, maret05}, where emission arises from a compact region ($<$1\arcsec, i.e., $<$250 AU) centred on the source \citep{jorgensen07}, and the same is likely true for the medium H$_2$O component in that source. In interferometric observations of IRAS4A, a small ($\sim$few arcsec) blue-shifted outflow knot of similar width has been identified in, e.g., SiO and SO \citep{choi05, jorgensen07}. Small-scale structures exist in the other sources as well, which may produce the medium components.

The H$_2$O 2$_{02}$--1$_{11}$ spectrum of IRAS4A shows an inverse P Cygni profile, a clear sign of infall also detected in other molecular tracers using interferometer observations \citep{difrancesco01, jorgensen07}. This infall signature is also tentatively seen in the 1$_{11}$--0$_{00}$ line, but here the absorption from the outer envelope dominates and little is left of the blue emission peak. The signature is not seen in higher-excitation lines. The separation of the emission and absorption peaks is $\sim$0.8 km\,s$^{-1}$, whereas it is $\sim$1.5 km\,s$^{-1}$ in the observations of \citet{difrancesco01} and larger in the observations by \citet{jorgensen07}, indicating that the infall observed in H$_2$O 2$_{02}$--1$_{11}$ takes place over larger spatial scales.

The passively heated envelope is seen in ground-based observations of high-density tracers to produce narrow emission, $<$3 km\,s$^{-1}$, which may be self-absorbed (Fig. \ref{fig:i2a_co65}). For water, this type of emission is not seen in any of the sources; the medium component is broader by a factor of 2--3 with respect to what is expected from the envelope. The absorption seen in all three sources is attributed to cold gas in the outer parts of the envelope. Using interferometric observations, \citet{Jorgensen10} detected compact, narrow ($\sim$1 km\,s$^{-1}$) emission in the H$^{18}_2$O 3$_{13}$--2$_{20}$ line in IRAS4B possibly originating in the circumstellar disk. Scaling the observed emission to the transitions observed here by assuming $T_{\rm rot}$=170 K \citep{Watson07}, the expected emission is typically less than 10\% of the rms for any given transition. Hence, when extrapolated to the disks surrounding IRAS2A and 4A, the disk contribution to the H$_2$O emission probed by HIFI is negligible. The H$_2$O excitation temperature of the broad component is 220$\pm$30 K, comparable to that found by \citet{Watson07}, but the inferred column density is a factor of 100 higher. Thus, the mid-infrared lines seen by \citeauthor{Watson07} may come from the same broad outflowing gas found by HIFI, provided the mid-infrared lines experience a factor of 100 more extinction.

\subsection{Abundances}\label{sect:abun}

\subsubsection{Shocks: H$_2$O/CO}

The observed broad components are compared directly with HIFI observations of CO 10--9 \citep{Yildiz10}, because the width and position of the lines are similar and they were obtained using approximately the same beamsize (22\arcsec\ versus 19\arcsec). The exception is for IRAS2A, where the blue line wing is not observed. The advantage is that no detailed models are required to account for the H$_2$O/CO abundance, as long as the lines are optically thin, in particular the emission from the wings. The abundance ratio is estimated for various temperatures by using the RADEX escape probability code \citep{vandertak07}. The density is assumed to be 10$^{5}$ cm$^{-3}$, appropriate for the large-scale core. If the emission is optically thin, the abundance ratio scales linearly with density resulting in the same line ratio corresponding to a higher abundance ratio. There is little variation in the predicted ratio for $T$$>$150 K, the typical temperature inferred by \citet{Yildiz10}. The line ratios and abundance ratios are listed as a function of velocity in Table \ref{tab:coh2o} in the online appendix.

The abundance ratio increases with increasing velocity from H$_2$O/CO of $\sim$0.2 near the line centre to H$_2$O/CO$\gtrsim$1 in the line wings of all sources for velocity offsets larger than 15 km\,s$^{-1}$ with respect to that of the source (Fig. \ref{fig:i2a_co65}). Assuming that the CO abundance is 10$^{-4}$, the H$_2$O abundances are in the range of $\sim$10$^{-5}$--10$^{-4}$. Only at high velocities is the temperature high enough for oxygen to be driven into water by means of the neutral-neutral reaction O + H$_2$ $\rightarrow$ OH + H; OH + H$_2$ $\rightarrow$ H$_2$O. The same result was found in the massive outflow in Orion-KL \citep{franklin08}, where less than 1\% of the gas in the outflow experiences this high-temperature phase. The fraction of gas for which the H$_2$O/CO abundance is $>$1 is $\sim$5--10\% for the sources observed here.

For IRAS2A, a deep spectrum of CO 6--5 obtained with CHAMP$^+$ on APEX simultaneously with observations of HDO 1$_{11}$--0$_{00}$ (Liu et al. in prep.) shows the same morphology in terms of a broad and medium component (Fig. \ref{fig:i2a_co65}). Furthermore, the velocity offset and FWHM are the same as for H$_2$O suggesting that the line profiles are not unique to H$_2$O, although the broad component is far more prominent in H$_2$O. The ratio of peak intensities for the two components is $\sim$2--3 in H$_2$O versus 10 in CO 6--5. Analysing the abundance ratio as a function of temperature shows that H$_2$O/CO$\sim$0.1--1 for $T$$>$150 K (Table \ref{tab:coh2o}), consistent with what is found for CO 10--9.

\subsubsection{Envelope}

The simplest way to constrain the H$_2$O abundance in the outer envelope is with calculations using RADEX on the narrow absorption in the 1$_{11}$--0$_{00}$ line. The absorption is optically thick -- in particular for IRAS2A and 4A, where the feature is saturated -- which requires a para-H$_2$O column density of $>$10$^{13}$ cm$^{-2}$ if one assumes typical values for $T$ and $n({\rm H}_2)$ of 15 K and 10$^5$ cm$^{-3}$. With a pencil-beam H$_2$ column of $\sim$$10^{23}$ cm$^{-2}$ \citep{jorgensen02}, the total H$_2$O abundance in the outer envelope is $\gtrsim$10$^{-10}$.

For IRAS2A and 4A, the H$_2$O abundance was further constrained using radiative transfer models. The setup is a spherical envelope with density and temperature profiles constrained from continuum data \citep{jorgensen09}, an infall velocity profile $\varv=(2\,{\rm km}\,{\rm s}^{-1})(r/r_{\rm in})^{-1/2}$, and a Doppler parameter $b=0.8$ km\,s$^{-1}$. Line fluxes were computed with the new radiative transfer code LIME (Brinch \& Hogerheijde subm.). The models constrained the abundance of water in the outer envelope to be $\sim$10$^{-8}$. Lower values are insufficient to obtain saturated absorption in the 1$_{11}$--0$_{00}$ line, and $\sim$10$^{-8}$ is the highest abundance where the resulting narrow emission can be hidden in the observed higher-excitation H$_2$O lines. The models predict that the H$_2$O emission from the warm inner envelope ($r$$\lesssim$100 AU) is optically thick, hence no constraints can be obtained from the H$_2$O spectra on the inner abundance. However, the lack of narrow H$_2^{18}$O emission infers an upper limit on the H$_2$O abundance of $\sim$10$^{-5}$ (Visser et al. in prep.).

\section{Conclusions}

These observations represent one of the first steps towards understanding the formation and excitation of water in low-mass star-forming regions by means of resolved line profiles. The three sources have remarkably similar line profiles. Both the H$_2^{16}$O and H$_2^{18}$O lines are very broad, indicating that the bulk of the emission originates in shocked gas. The broad emission also highlights that water is a  far more reliable dynamical tracer than, e.g., CO. Comparing C$^{18}$O to H$^{18}_2$O emission and line profiles indicates that the H$_2$O/CO abundance is high in outflows and low in the envelope. Additional modelling of the emission, should be able to constrain the total amount of water in the envelope and outflowing gas, thus test the high-temperature gas-phase chemistry models for the origin of water. This will be performed for a total sample of the 29 low-mass YSOs to be observed within the WISH key programme.

\onltab{3}{
\begin{table}
\caption{CO 6--5 and CO 10--9/H$_2$O 2$_{02}$--1$_{11}$ line ratios in 5 km\,s$^{-1}$ intervals and corresponding abundance ratio for $T$$>$150 K and $n$=10$^{5}$ cm$^{-3}$.}
\scriptsize
\begin{center}
\begin{tabular}{r c c c c c c c c}
\hline \hline
d$\varv_{\rm LSR}$ & \multicolumn{4}{c}{IRAS2A} & \multicolumn{2}{c}{IRAS4A}  & \multicolumn{2}{c}{IRAS4B} \\
(km\,s$^{-1}$) & CO 6--5/ & $x$(H$_2$O)/ & CO 10--9/ & $x$(H$_2$O)/ & CO 10--9/ & $x$(H$_2$O)/ & CO 10--9/ & $x$(H$_2$O)/ \\
 & H$_2$O 2$_{02}$--1$_{11}$ & $x$(CO) & H$_2$O 2$_{02}$--1$_{11}$ & $x$(CO) & H$_2$O 2$_{02}$--1$_{11}$ & $x$(CO) & H$_2$O 2$_{02}$--1$_{11}$ & $x$(CO) \\
\hline
     $-$20 --  $-$15 &   5.0 &  0.34 & \ldots &  \ldots &   0.8 & 1.11 & \ldots & \ldots \\
     $-$15 --  $-$10 &   3.8 &  0.45 & \ldots &  \ldots &   2.0 & 0.43 &   0.9 &  1.00  \\
     $-$10 --   $-$5 &   4.6 &  0.37 & \ldots &  \ldots &   2.8 & 0.31 &   1.4 &  0.64  \\
      $-$5 --      0 &   9.3 &  0.18 & \ldots &  \ldots &   2.4 & 0.36 &   1.7 &  0.50  \\
       0 --        5 &  26.6 &  0.06 & \ldots &  \ldots &   2.9 & 0.29 &   2.3 &  0.37  \\
       5 --       10 &  17.0 &  0.10 &    3.4 &    0.26 &   3.9 & 0.22 &   2.9 &  0.29  \\
      10 --       15 &  11.1 &  0.15 &    3.2 &    0.27 &   3.3 & 0.26 &   2.1 &  0.42  \\
      15 --       20 &   3.5 &  0.48 &    0.9 &    1.00 &   2.4 & 0.36 &   1.6 &  0.53  \\
      20 --       25 &   1.4 &  1.25 &    0.4 &    2.00 &   1.0 & 0.83 &   1.1 &  0.77  \\
      25 --       30 & \ldots & \ldots & \ldots & \ldots &  0.8 & 1.11 &   0.5 &  1.67  \\
      30 --       35 &   0.9 &  2.00 & \ldots & \ldots & \ldots & \ldots & 0.9 &  1.00  \\
\hline
\end{tabular}
\end{center}
\label{tab:coh2o}
\end{table}
}

\bibliographystyle{aa}
\bibliography{bibliography}

\begin{acknowledgements}
This work is made possible thanks to the HIFI guaranteed time programme. HIFI has been designed and built by a consortium of institutes and university departments from across Europe, Canada and the US under the leadership of SRON Netherlands Institute for Space Research, Groningen, The Netherlands with major contributions from Germany, France and the US. Consortium members are: Canada: CSA, U.Waterloo; France: CESR, LAB, LERMA, IRAM; Germany: KOSMA, MPIfR, MPS; Ireland, NUI Maynooth; Italy: ASI, IFSI-INAF, Arcetri-INAF; Netherlands: SRON, TUD; Poland: CAMK, CBK; Spain: Observatorio Astronomico Nacional (IGN), Centro de Astrobiolog{\'i}a (CSIC-INTA); Sweden: Chalmers University of Technology - MC2, RSS \& GARD, Onsala Space Observatory, Swedish National Space Board, Stockholm University - Stockholm Observatory; Switzerland: ETH Z{\"u}rich, FHNW; USA: Caltech, JPL, NHSC. HIPE is a joint development by the Herschel Science Ground Segment Consortium, consisting of ESA, the NASA Herschel Science Center, and the HIFI, PACS and SPIRE consortia. We thank many funding agencies for financial support.
\end{acknowledgements}

\end{document}